\documentclass[12pt,twoside]{article}
\usepackage{amssymb}
\usepackage{amsmath}
\usepackage{latexsym}
\usepackage{longtable}
\usepackage{epsfig}
\usepackage{eucal}
\usepackage{color}
\usepackage{graphicx}
\usepackage{amsfonts} 
\usepackage{amsmath}
\usepackage{amssymb}
\def\bn{{\bf{n}}}
\def\be{{\bf{e}}}

\begin{document}
\vspace*{2cm}
\begin{center}
 \LARGE{Fourier's law based on microscopic dynamics}\bigskip
 \end{center}
\begin{center}
\large{Abhishek Dhar$^1$ and Herbert Spohn$^2$}
 \end{center}
 \begin{center}
{ $^1$International Centre for Theoretical Sciences - Tata Institute of Fundamental Research$,$ Bengaluru 560089 $,$ India \\$^2$ Zentrum Mathematik and Physik Department, Technische Universit\"{a}t M\"{u}nchen, Garching, Germany} \\
\tt{abhishek.dhar@icts.res.in}, \tt{spohn@tum.de}
 \end{center}
\vspace*{4cm}
\textbf{Abstract}. While Fourier's law is empirically confirmed for many substances and over an extremely wide range of 
thermodynamic parameters, a convincing microscopic derivation still poses difficulties. With current machines the solution of 
Newton's equations of motion can be obtained with high precision and for a reasonably large number of particles.
For simplified model systems one thereby arrives at a deeper understanding of the microscopic basis for Fourier's law.
We report on recent, and not so recent, advances. 
\vspace*{3cm}
\begin{flushright} 5.4.2019
\end{flushright}

\newpage
 \section{Introduction}
\label{sec1}
In its modern form the law of heat conduction consists of two parts. The first part states the conservation of energy in a continuum form as
\begin{equation}\label{1.1}
\partial_t e + \nabla_x\cdot j_e = 0.
\end{equation} 
Here $e(x,t)$ is the local energy at location $x$ and time $t$, while $j_e(x,t)$ is the local energy current. $x\in\mathbb{R}^3$ would be physical space, but later on we also consider models in one dimension for which  $x \in \mathbb{R}$, $ \nabla_x = \partial_x$. The second part, Fourier's law proper, states that the energy current is proportional to the gradient of the local temperature, $T(x,t)$,
\begin{equation}\label{1.2}
j_e = -\kappa \nabla_x T 
\end{equation} 
with $\kappa$ the conductivity matrix.  Introducing the heat capacity $c = \partial e/\partial T$, equivalently   
\begin{equation}\label{1.2a}
j_e =-\frac{\kappa}{c} \nabla_x e
\end{equation} 
 and, combining with \eqref{1.1}, one arrives at
\begin{equation}\label{1.3}
\partial_t e  = \nabla_x\cdot \frac{\kappa}{c} \nabla_x e.
\end{equation} 
If $\kappa$ and $c$ are independent of $e$, then \eqref{1.3} is a linear equation which, in infinite space, can be solved through Fourier transform as
\begin{equation}\label{1.4}
\partial_t \hat{e}(k,t)   = -\frac{(k\cdot \kappa k)}{c} \hat{e}(k,t).
\end{equation} 

Experimentally, the local temperature, $T(x,t)$, is more easily accessible. Then in \eqref{1.3} $e(x,t)$   is substituted by  $T(x,t)$.
The thermal conductivity is tabulated for a wide range of substances. There is no doubt that empirically Fourier's law is of an essentially universal  applicability. 
 Fourier's \textit{Th\'{e}orie analytique de la chaleur} was published in 1822 \cite{fourier}. A few decades later, physical matter was conceived as made up of many molecules with their motion
 governed by Newton's equation of motion. Thus the issue  of deriving Fourier's law from an underlying  microscopic theory was moved to the agenda. 
 In an early triumph, Boltzmann and Maxwell  used kinetic theory to predict the thermal conductivity of gases. Peierls extended the kinetic theory to  solids and Nordheim
 to weakly interacting electrons. Still nowadays kinetic theory is used as a basic tool for computing conductivities. But for strongly interacting systems, with no obvious small parameter at hand, the problem remains challenging. 
 
 There are some obvious constraints for \eqref{1.2} to be valid.  Fourier's law refers to a single conservation law. Generically, one has to expect several conservation laws, which in principle requires to handle all of them on equal footing. Secondly, $\kappa$ will itself depend on energy and the diffusion equation
 \eqref{1.3} becomes nonlinear. 
  
 In a mechanical model, for which the motion of molecules is  governed by a Hamiltonian with a short range potential, energy is locally conserved. Therefore local energy conservation holds for every history, which in essence implies the validity  \eqref{1.1} on a macroscopic scale  for typical initial data. The true difficulty arises in establishing Fourier's law
 for a specific model Hamiltonian \cite{bonetto2000}. Just to mention a single item, in \eqref{1.2} the response of the current to a temperature gradient is modeled as strictly local. However, microscopically the response will take some time and will be smeared over a spatial region. Thus one has to invoke dynamical properties of the motion of many molecules which ensure the validity of \eqref{1.2} on a suitably coarse space-time scale.
 
   \section{Defining Fourier's law}
\label{sec1}
We proceed with a specific model, describing a one-dimensional solid with anharmonic couplings. But the method easily extends to more complicated models.
We consider the one-dimensional lattice $\mathbb{Z}$, or possibly some segment $[N_1,...,N_2]$. The atoms have displacements $q_j$ and the associated momentum
$p_j$, $j\in  \mathbb{Z}$. Without restriction we set the mass equal to $1$. The motion of the atoms is confined by an on-site potential $U$. In addition the atoms are coupled through the potential $V$.
Both potentials are bounded from below and $U(q), V(q) \to \infty$ as $|q|\to \infty$. Then our model Hamiltonian reads
\begin{equation}\label{2.1}
H = \sum_{j \in \mathbb{Z}}\big(\tfrac{1}{2} p_j^2 + U(q_j) + V(q_{j+1} - q_j)\big).
\end{equation} 
 The equations of motion are
\begin{equation}\label{2.2}
\ddot{q}_j = -U'(q_j)  +V'(q_{j+1} - q_j) - V'(q_{j} - q_{j-1}).
\end{equation} 
For a generic choice of $U,V$ the energy is expected to be the only locally conserved field. The local energy is 
\begin{equation}\label{2.3}
e_j = \tfrac{1}{2} p_j^2 + U(q_j) + V(q_{j+1} - q_j)
\end{equation} 
and from \eqref{2.2} we infer the local energy current
\begin{equation}\label{2.4}
J_j = -p_j V'(q_{j} - q_{j-1})
\end{equation} 
with the property that
\begin{equation}\label{2.5}
\dot{e}_j -J_{j-1} +J_j = 0,
\end{equation} 
obviously the discrete version of \eqref{1.1}. 

There are two conceptually different ways to define Fourier's law. The first one focuses on heat diffusion. For this purpose we introduce the Gibbs distribution
\begin{equation}\label{2.6}
(Z_N)^{-1} \exp[-\beta H_N] \prod_{j=1}^N \mathrm{d}p_j \mathrm{d}q_j
\end{equation} 
with $\beta = 1/k_\mathrm{B}T$ the inverse temperature and $H_N$ as in \eqref{2.1}, but the sum restricted to $j = 1,...,N-1$. We will use 
temperature units such that $k_\mathrm{B}= 1$. Since the model is in one dimension, the infinite volume
limit is well-defined and the limit state has a finite correlation length, of course depending on $\beta$. Averages at infinite volume are denoted by
$\langle \cdot \rangle_\beta$ and  the connected correlation by $\langle \cdot \rangle_\beta^\mathrm{c}$. We now introduce the energy time-correlation through 
\begin{equation}\label{2.7}
C_{ee}(j,t) = \langle e_j(t)e_0(0) \rangle_\beta^\mathrm{c},
\end{equation} 
which can be viewed as perturbing the thermal state at the origin and measuring how the surplus energy propagates. Since $j,t \mapsto e_j(t)$ is stationary
under space-time translations, to fix one argument at $j=0,t=0$ is just a standard convention. By space-time reversal, $C_{ee}(j,t) = C_{ee}(-j,t) = C_{ee}(j,-t)$. 
The static energy susceptibility is defined through
\begin{equation}\label{2.7a}
\chi_{ee} = \sum_{j \in \mathbb{Z}} C_{ee}(j,t) = \sum_{j \in \mathbb{Z} }C_{ee}(j,0),
\end{equation} 
where the second identity follows from the conservation law.
For small $|j|,|t|$ the energy correlator depends on specific microscopic details. But considering larger arguments, $C_{ee}(j,t)$ should be well approximated by the heat diffusion kernel
\begin{equation}\label{2.8}
\chi_{ee}(4\pi D|t|)^{-\frac{1}{2}} \exp[-j^2/4D |t|],
\end{equation} 
the prefactor ensuring the normalization \eqref{2.7a}. In particular we have achieved a microscopic definition of the energy diffusivity $D = D(\beta)$. 
By taking second moments in \eqref{2.7} and using \eqref{2.5}, one obtains a Green-Kubo identity for $D(\beta)$ as  
  \begin{equation}\label{2.9}
D(\beta) = \frac{1}{\chi_{ee}}\int_0^\infty \mathrm{d}t C_J(t), \qquad C_J(t) = \sum_{j \in \mathbb{Z}} \langle J_j(t)J_0(0) \rangle_\beta.
\end{equation} 
No truncation is needed, since $\langle J_0(0) \rangle_\beta = 0$. 

In general, $D$ depends on $\beta$ and thereby on $e$ through the thermodynamic relation
$e(\beta) =  \langle  e_0\rangle_\beta$. Thus heat will propagate according to a nonlinear diffusion equation, as already argued before. However, by our choice of initial conditions, we have to 
match the microscopic evolution with the linearized diffusion equation. The nonlinearity would become manifest by imposing initially a slowly varying, non-constant energy profile.

The Green-Kubo formula well illustrates the difficulties.
The chain is strongly interacting, except when both potentials are harmonic. Small deviations from harmonicity could be handled by kinetic theory. But otherwise the dynamics is chaotic. No analytic theory should be expected. Fortunately these days we have the capability to run molecular dynamics (MD) simulations on large systems with high precision.  While our model is strongly simplified,  MD teaches us a lot. One can check how fast
the ideal theoretical value of $D$ is approached, if at all, how $D$ depends on $\beta$, and many more details.
In fact, when by modern standards the still fairly primitive computers became available to physicists, one of the first MD simulations studied 300 hard disks in a quadratic volume 
with a density of $\tfrac{1}{2}$ relative to close packing \cite{alder1970}. 
To the then big surprise,  the total momentum current time-correlation was discovered to decay only as $|t|^{-1}$, which 
triggered the theoretical investigations on long-time tails of current time-correlations. 
  
The second approach focuses directly on Fourier's law through the energy current induced by a small temperature gradient \cite{LLP2003,dhar2008}, which  is closer to an experimental set up. The basic idea is
to impose thermal boundary conditions  on \eqref{2.2}. Theoretically preferred are Langevin thermostats. In some MD simulations 
one adds sufficiently chaotic deterministic reservoir dynamics, as Nose-Hoover thermostats, which are designed so to act as a heat bath. Thereby
a somewhat faster algorithm is achieved, which is however less significant with modern machines.
We choose $j = 1$ and $j = N$ as boundary points. In addition to the deterministic motion \eqref{2.2} we modify the dynamics of  the first site to
\begin{equation}\label{2.10}
\dot{q}_1(t) = -U'(q_1(t)) - V'(q_1-q_0)+V'(q_2-q_1) -\gamma p_1(t) + \sqrt{2\gamma T_{\mathrm{L}}} \xi(t)~, 
\end{equation} 
where the extra variable $q_0$ is introduced to fix boundary conditions (e.g $q_0=0$ for fixed boundary conditions), and 
 $\xi(t)$ is standard Gaussian white noise with correlator $\langle\xi(t) \xi(t')\rangle = \delta(t - t')$. The friction coefficient $\gamma$ is a free parameter.
$T_{\mathrm{L}}$ is the temperature of the left reservoir. If there is only coupling to this heat bath, the other end being left free, then in the long time limit the system approaches thermal equilibrium at temperature $T_{\mathrm{L}}$. Correspondingly we add a reservoir at $j = N$ with temperature
 $T_{\mathrm{R}}$. If $T_{\mathrm{L}}= T_{\mathrm{R}}$, again thermal equilibrium is the unique steady state. But if $T_{\mathrm{L}} \neq T_{\mathrm{R}}$
 a nonequilibrium steady state (NESS) is imposed with no explicit formula for its probability density function at hand. In MD one would run the system for a long time until  in approximation statistical stationarity is observed. The average in the corresponding steady state is here denoted by
 $\langle\cdot\rangle_{T_\mathrm{L},T_\mathrm{R}}$. In fact, under suitable conditions on $U,V$ it is known that the steady state 
 has a smooth density and the stochastic generator has a spectral gap, implying exponentially fast convergence to stationarity \cite{eckmann2000}.
 While the mathematical proofs are ingenious, they cannot access the $N$-dependence which is the crux of Fourier's law. Because of the conservation law 
 the steady state current is constant in $j$ and, based on the solution of \eqref{1.4} with thermal boundary conditions, one expects that 
\begin{equation}\label{2.11}
\lim_{N\to \infty} N \langle J_j \rangle_{T_\mathrm{L},T_\mathrm{R}}  = \eta( T_\mathrm{L},T_\mathrm{R})
\end{equation} 
with $\eta$ depending on both boundary temperatures. In linear response, $T_\mathrm{L} = T +\tfrac{1}{2} \Delta T$,
$T_\mathrm{R} = T -\tfrac{1}{2} \Delta T$, one obtains
\begin{equation}\label{2.12}
\eta( T+\tfrac{1}{2} \Delta T,T - \tfrac{1}{2} \Delta T) \simeq \kappa \Delta T.
\end{equation} 
The thermal conductivity, $\kappa$, depends on $T$ in general. The linear response argument can be carried out also for the microscopic model, resulting in
\begin{equation}\label{2.13}
N \langle J_j \rangle_{T +\frac{1}{2} \Delta T,T -\frac{1}{2} \Delta T}  
= T^{-2} \Delta T \int_0^\infty \mathrm{d}t \sum_{j \in \mathbb{Z}} \langle J_j(t)J_0(0) \rangle_\beta
\end{equation}  
for large $N$, thereby confirming the Einstein relation
\begin{equation}\label{2.14}
\kappa =\frac{\chi_{ee} }{T^2} D = cD,
\end{equation} 
since the specific heat satisfies $c=\chi_{ee}/T^2$ according to thermodynamics.
\section{Molecular dynamics of model systems}

We discuss three examples of Hamiltonian systems for which the validity of Fourier's law one can tested through numerical simulations. The examples include two one-dimensional systems and one three-dimensional system. To measure transport coefficients by numerical means has a long tradition,
to mention only \cite{alder21970,casati1984,lepri1997}. To obtain finer details, as full energy and energy current distributions and steady state profiles, is 
more recent and obviously reflects the increased computational power.
\subsection{One-dimensional $\phi^4$ chain}

\begin{figure}
 \includegraphics[width=4.0in]{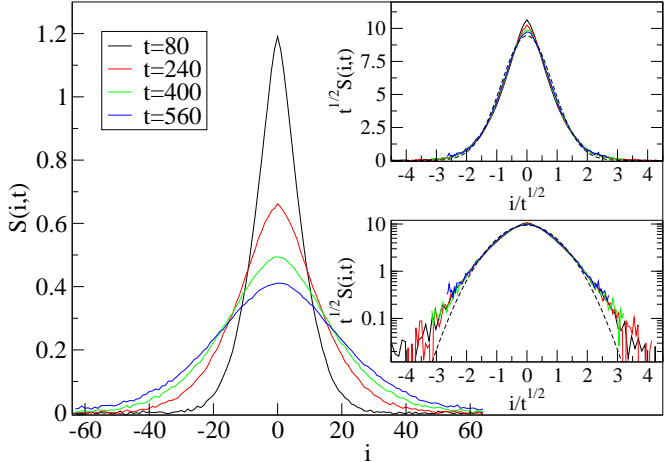}
\caption{$\phi^4$ chain: Plot of correlations in equilibrium energy fluctuations $C(i,t)$ for a system of size $N=128$ at different times. The top inset show a diffusive scaling of the data while the lower inset shows the same on a logarithmic scale. The dashed-line indicates a Gaussian fit given by $f(x,t)=\chi_{ee} e^{-i^2/(4 D t)}/\sqrt{4 \pi D t}$, with $\chi_{ee}=\sum_i C(i,0)=20.36$ and $D=0.425$. }\label{fig:1}
\end{figure}
\begin{figure}
 \includegraphics[width=4.in]{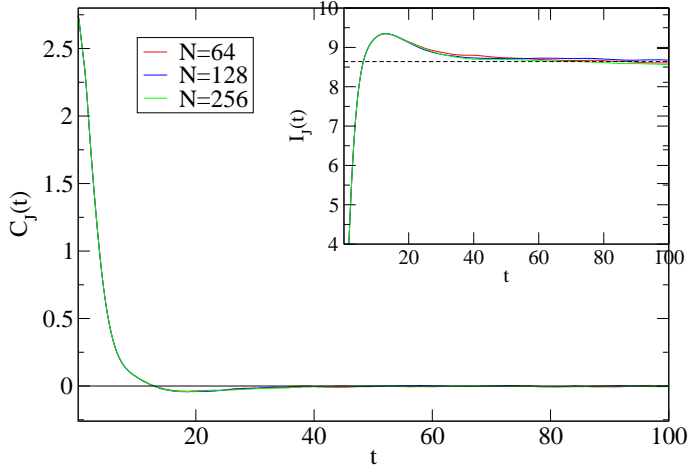}
\caption{$\phi^4$ chain: Plot of the current auto-correlation function $C_J(t)= \sum_i \langle J_i(t) J_0(0) \rangle$, for three system sizes. The inset shows the integral  $I_J(t)=\int_0^t ds C_J(s)$ which can be seen to saturate to around the value $I=8.64$ (dashed line) at large times. }
\label{fig:2}
\end{figure}
\begin{figure}
 \includegraphics[width=4.in]{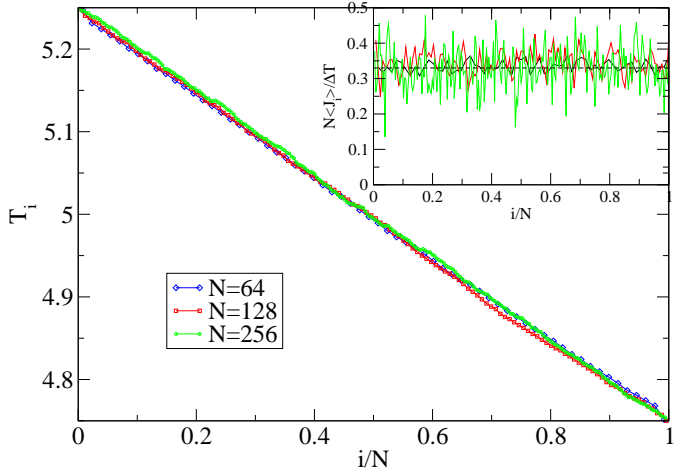}
\caption{$\phi^4$ chain: Plot of temperature profile in the NESS for different system sizes. The boundary temperatures were taken as $T_\mathrm{L} =5.25$ and $T_\mathrm{R}=4.75$. The inset shows the scaled local heat current profile. The dashed line is the estimated thermal conductivity $\kappa=0.337$.}
\label{fig:3}
\end{figure}
We consider the Hamiltonian \eqref{2.1} with the choices $U(q)=q^4/4$, $V(r)=r^2/2$. Nonequilibrium as well as equilibrium simulations of this model were performed in \cite{Zhao00,Aoki02,Zhao06} and clearly demonstrated the validity of Fourier's law. A kinetic theory treatment was presented in \cite{Aoki06,Lefevere06} and a formula for the conductivity, valid in the weak-nonlinearity limit was proposed and tested in simulations. Here we present results from recent both nonequilibrium as well as equilibrium simulations. \medskip\\
\emph{Equilibrium simulations}. The sum in Eq.~\eqref{2.1} is restricted to $j=1,2,\dots,N$ and with periodic boundary conditions $q_{N+1}=q_1$. In Fig.~\eqref{fig:1} we show plots for the energy correlator in equilibrium as defined in \eqref{2.7}. The data have been generated by averaging over $4\times 10^5$ initial conditions chosen from a Gibbs distribution at temperature $T=5.0$.  We observe a diffusive spreading of the correlations, following the prediction of Eq.~\eqref{2.8}, with a diffusion constant $D \approx 0.425$.   
As a test of the Green-Kubo identity \eqref{2.9}, we also evaluated the energy current auto-correlation $C_J(t)$. The data are shown in Fig.~(\ref{fig:2}), where the time-integral is seen to converge to a value $I\approx 8.64$. 
According to \eqref{2.9}, one expects  $I= \chi_{ee}D$. Indeed this is  consistent with our numerical results, 
since $\chi_{ee}D   =   20.36\times  0.425=8.653$, while $I=8.64$.\medskip\\
\emph{Nonequilibrium simulations}. We  consider a finite chain of $N$ particles $i=1,2,\ldots,N$ and consider the Hamiltonian \eqref{2.1} with tied down boundary conditions which means to add the extra interaction terms $(q_1-q_0)^2/2$, resp. $(q_{N+1}-q_N)^2/2$, and setting $q_0=0 =q_{N+1}$.  The particles $i=1$ and $i=N$ are connected to Langevin heat baths as in \eqref{2.10}. We  choose the left and right bath temperatures to be $T_\mathrm{L}=5.25$ and $T_\mathrm{R}=4.75$ respectively, so that the mean temperature $T=(T_\mathrm{L}+T_\mathrm{R})/2=5.0$ is the same as that in the equilibrium simulations, and the temperature difference $\Delta T=T_L-T_R=0.5=0.1 T$ is relatively small, thus still in the linear response regime. In Fig.~(\ref{fig:3}) we plot the temperature profile and the local current profile obtained in the nonequilibrium steady state for system sizes $N=64,128,256$. The linear temperature profile suggests that the thermal conductivity does not vary much in the given small temperature range and we expect that the thermal conductivity is obtained through $\kappa= N \langle J_i
\rangle /\Delta T$. From our data for different sizes, we estimate $\kappa=0.337$.  We confirm also the Einstein relation \eqref{2.14}, since
our numerical results yield  for the right-hand-side $(20.36/25) \times 0.425=0.34442$, to be compared with $\kappa=0.337$ from nonequilibrium simulations.

\subsection{One-dimensional Heisenberg spin chain}
Spin models provide an interesting class of models to study transport. Several studies on the rotor chain (equivalent to the $XY$ model) have verified Fourier's law in this system at finite temperatures \cite{Gendelman2000,Das2015,Iubini2016}. Here we report studies  of the Heisenberg spin chain which is a well-known model in condensed matter physics and widely used to understand magnetic systems. 
The properties of dynamical correlations in this system were recently studied \cite{das2019} and we report on some results relevant to our discussion here.
The spins are of unit length and arranged along a one-dimensional lattice, $\vec{S}_j = (S_j^x,S_j^y,S_j^z)$ with $| \vec{S}_j| = 1$, $j=1,2,\ldots N$. The system is described by the Hamiltonian
\begin{equation}\label{3.2}
H = - \sum_{j =1}^N \big( S_j^xS_{j+1}^x + S_j^yS_{j+1}^y + \Delta S_j^zS_{j+1}^z\big)
\end{equation}
and we consider periodic boundary conditions $\vec{S}_{N+1}=\vec{S}_1$.
The equations of motion are 
\begin{equation}\label{3.3}
\tfrac{d}{dt}\vec{S}_j = \{\vec{S}_j,H\}= \vec{S}_j \times \vec{B}_j,\quad \vec{B}_j = -\nabla_{\vec{S}_j} H,
\end{equation}
where the Poisson bracket between two functions, $g_1,g_2$, of the spin variables is defined by $\{g_1,g_2 \} = \sum_j\epsilon_{\alpha\beta\gamma} (\partial g_1/\partial {S_j^\alpha}
) (\partial g_2/\partial {S_j^\beta}) S_j^\gamma$ with the usual summation convention.  To see the Hamiltonian character of the dynamics it is instructive to  introducing the position-like angular variable $\phi_j \in S^1$ and the conjugate canonical momentum-like variable $s_j \in [-1,1]$
defined through
\begin{equation}\label{3.4}
S_j^x = f(s_j) \cos \phi_j,\quad S_j^y = f(s_j) \sin \phi_j,\quad S_j^z = s_j ,
\end{equation}
where $f(x) =(1-x^2)^{1/2}$.  In these variables the Hamiltonian \eqref{3.2}  reads
\begin{equation}\label{3.5}
H = - \sum_{j \in \mathbb{Z}} \Big( f(s_j) f(s_{j+1}) \cos(\phi_j- \phi_{j+1}) + \Delta s_js_{j+1} \Big),
\end{equation}
while the equations of motion take the standard Hamiltonian form
\begin{align}\label{3.6}
\tfrac{d}{dt}\phi_j = \frac{\partial H}{\partial s_j}, ~~
\tfrac{d}{dt}s_j = -\frac{\partial H}{\partial \phi_j}. 
\end{align}
This system is expected to be non-integrable with two conserved fields only, namely the total energy $H$ and the total $z$-magnetization $M=\sum_j S^z_j$. Hence, at any finite temperature,  it is natural to consider the equilibrium distribution $e^{-\beta(H - h M)}/Z(\beta,h)$, where $h$ is an external  magnetic field. 
It can be shown that the corresponding currents have vanishing equilibrium expectations. 
\begin{figure}
 \includegraphics[width=4.in]{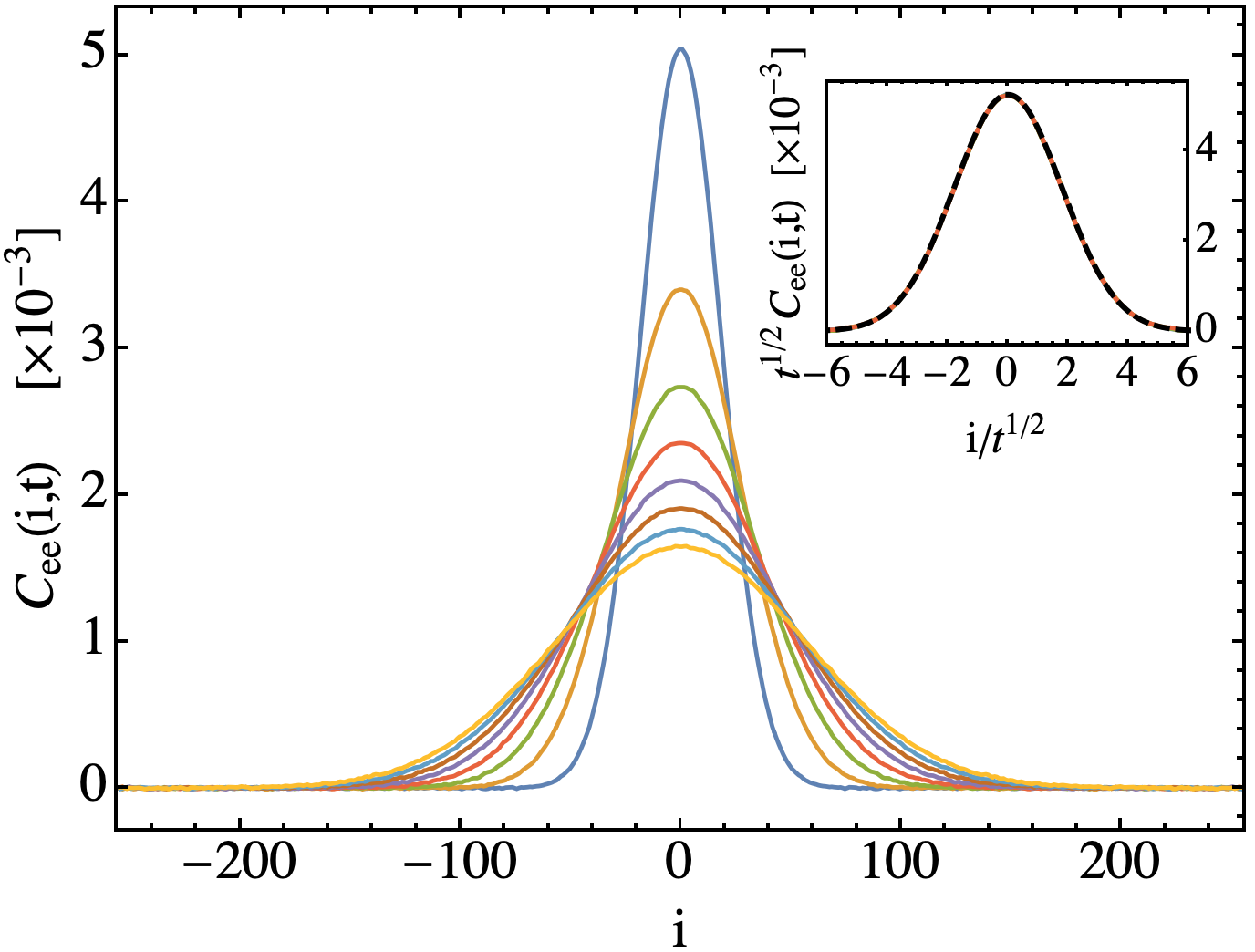}
\caption{Heisenberg spin chain: plot of 
energy correlations $C_{ee}(i,t)$ at different times. The inset shows collapse of the data under diffusive scaling. The parameter values for this simulation were $\Delta=1/2, \beta=1, h=0, N=512$. (Reprinted from \cite{das2019})}
\label{fig:4}
\end{figure}
\begin{figure}
 \includegraphics[width=4.in]{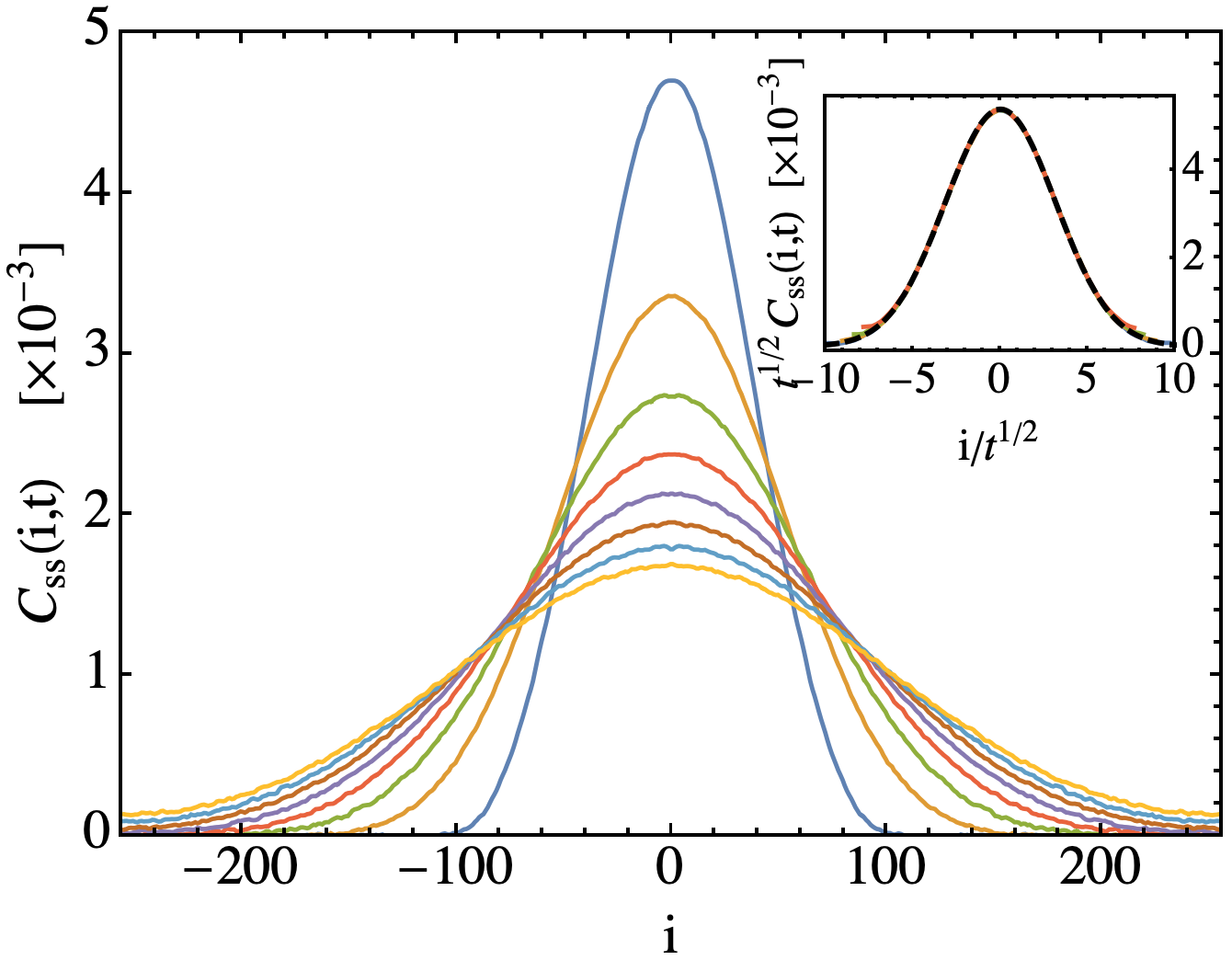}
\caption{Heisenberg spin chain: plot of 
$S^z$ correlations $C_{ss}(i,t)$ at different times. The inset shows collapse of the data under diffusive scaling. The parameter values for this simulation were $\Delta=1/2, \beta=1, h=0, N=512$. (Reprinted from \cite{das2019})}
\label{fig:5}
\end{figure}

On the theory side one has to extend the discussion in Sect. 2 to several conserved fields. Thus $D,\chi,\kappa$ become $n\times n $ matrices, $n=2$ for the XXZ chain. The susceptibility $\chi$ is a symmetric matrix. To ensure stability $D$ must have positive eigenvalues, but is not symmetric, in general. Onsager realized that the microscopic total current matrix, compare with \eqref{2.9},  is symmetric by definition. For the XXZ model, spin and energy have opposite signs under time-reversal. Thus the  current cross-correlation vanishes and $D$ is diagonal.
Specifically we consider the equilibrium dynamical correlators 
\begin{align}\label{3.7}
C_{ee}(j,t)&=\langle e_j(t) e_0(t) \rangle^c_{\beta,h}~, \\
C_{ss}(j,t)&=\langle s_j(t) s_0(t) \rangle^c_{\beta,h}~,
\end{align} 
where the energy density is defined as $ e_j=S_j^xS_{j+1}^x + S_j^yS_{j+1}^y + \Delta S_j^zS_{j+1}^z$. In Figs.~(\ref{fig:4},\ref{fig:5}), simulation results 
are shown for $C_{ee}(j,t)$, $C_{ss}(j,t)$ at different times and for parameter values $\beta=1.0$ and $h=0.0$. The insets confirm that both correlations spread diffusively. The simulation results were obtained by averaging over $10^5$ equilibrium initial conditions. Also,  the cross correlations 
have  indeed a small amplitude \cite{das2019}.

\subsection{Fermi-Pasta-Ulam model in three dimensions}
As third example we discuss a three-dimensional momentum conserving system for which  simulation results indicate the anticipated validity of Fourier's law. 
We consider a $3D$ block of dimensions $N\times W \times W$,  with a scalar displacement field $q_\bn$ defined on  each lattice site $\bn=(n_1,n_2,n_3)$ where $n_1=1,2,...,N$ and $n_2=n_3=1,2,...,W$. The Hamiltonian is taken to be of the
Fermi-Pasta-Ulam (FPU) form,
\begin{equation}\label{3.8}
H=\sum_{\bn} \tfrac{1}{2} p_\bn^2+ \sum_{\bn,\be} \big(~\tfrac{1}{2} (q_\bn-q_{\bn+\be})^2
+ \tfrac{\nu}{4} (q_\bn-q_{\bn+\be})^4 \big),
\end{equation}
\begin{figure}
 \includegraphics[width=4.0in]{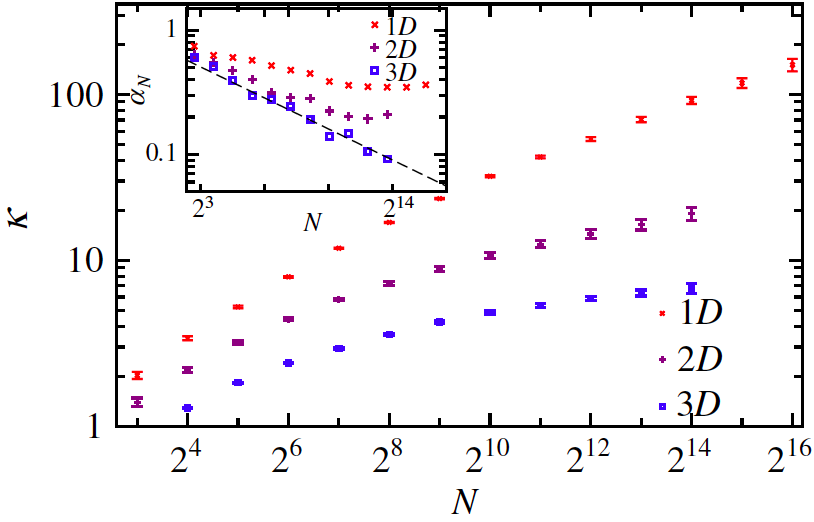}
\caption{Fermi-Pasta-Ulam model in different dimensions: plot of the effective thermal conductivity $\kappa=\langle J \rangle N/\Delta T$ as a function of length $N$, in different dimensions. The inset plots the running slope $\alpha_N = d \log \kappa /d \log N$, and in $3D$ this is clearly seen to decay with system size, while in lower dimensions they tend to saturate to finite values. The parameter values for this simulation were $T_\mathrm{L}=2, T_\mathrm{R}=1, \nu =2$. (Reprinted from \cite{saito2010})}
\label{fig:6}
\end{figure}
where $\be$ denotes a unit vector pointing along one of the three lattice axes. 
We have set the strength of the harmonic spring constants to
one. The anharmonicity parameter is $\nu$. For this model we will  discuss only results from nonequilibrium studies that were reported in \cite{saito2010}. To enforce a NESS, 
two of the faces of the crystal, namely those
at $n_1=1$ and $n_1=N$, are coupled to white noise Langevin heat baths. The equations of motion are then  given by
\begin{eqnarray}\label{3.9}
&&\hspace{-40pt}\ddot{q}_\bn=-\sum_\be \big((q_\bn-q_{\bn+\be}) + \nu (q_\bn-q_{\bn+\be})^3 \big)
\nonumber\\ 
&&\hspace{20pt}+  \delta_{n_1,1} (-\gamma \dot{q}_\bn+\sqrt{2 \gamma T_\mathrm{L} }\xi_{\bn}) +  \delta_{n_1,N}
(-\gamma \dot{q}_\bn+\sqrt{2 \gamma T_\mathrm{R} }\xi_{\bn}). 
\end{eqnarray} 
At a given site one has added normalized Gaussian white noise, compare with \eqref{2.10}, while  the noise terms are uncorrelated  at distinct sites. 
As before, tied down boundary conditions
are used for the particles connected to the baths and periodic
boundary conditions are imposed in the other directions.
We study heat current and temperature profile in the nonequilibrium steady state. The heat current $j_\bn$ from the lattice site
$\bn$ to $\bn+\be_1$ where $\be_1=(1,0,0)$,  is given by $ j_\bn =
F_{\bn,\bn+\be_1} \dot{q}_{\bn+\be_1} $, with $F_{\bn,\bn+\be_1}$ being the
force on the particle at site $\bn+\be_1$ due to the particle at site $\bn$. 
The steady state average current per bond is given by 
\begin{equation}\label{3.10}
\langle J \rangle_\mathrm{NESS}=\frac{1}{{W}^2(N-1)} \sum_{n_1=1}^{N-1} \sum_{n_2,n_3=1}^{W} \langle j_\bn \rangle_\mathrm{NESS},  
\end{equation}
while the average  temperature across  layers in the slab is given by 
\begin{equation}\label{3.10}
T_{n_1}=(1/{W}^2)\sum_{n_2,n_3} \langle\dot{q}_\bn^2 \rangle_\mathrm{NESS}. 
\end{equation}
In Fig.~(\ref{fig:6}) the size-dependence of the effective conductivity $\kappa = \langle J \rangle N/\Delta T$ is plotted as a function of $N$ for a sample of dimensions $N\times 32\times32$. For comparison, we also display corresponding results for one- and two-dimensional lattices. There is good evidence for a finite conductivity in $3D$, while it seems to diverge in lower dimensions.  

\section{Conclusions}
Note that the FPU lattice in one dimension is a special case of the Hamiltonian \eqref{2.1}. The on-site potential $U = 0$ and the coupling potential is anharmonic.
Our simulations suggest that the steady state energy current is of order $N^{-1 +\alpha}$ with some $\alpha > 0$, hence larger than predicted by Fourier's
law. So how come? In fact, super-diffusive transport in one dimension was first noted numerically in \cite{lepri1997} and later widely confirmed with $\alpha$ in the range
$0.3 - 0.5$ depending on the choice of $V$, see the  reviews and survey articles  \cite{LLP2003,dhar2008,Book}.

If $U \neq 0$, energy is the  only conservation law. But for the FPU chain, with general $V$, in addition momentum is conserved and also the free volume,
microscopically the stretch $r_j = q_{j+1} - q_j$. This by itself would not necessarily imply super-diffusive transport. But for $U=0$  the equilibrium states are of the form
\begin{equation}\label{4.1}
Z^{-1} \prod_{j=1}^N \exp\big(-\beta\big( \tfrac{1}{2}(p_j- u)^2 + V(r_j) + Pr_j\big)\big) 
\end{equation}
with $u$ the mean velocity and $P>0$ the pressure. Because of momentum conservation, there are equilibrium states with a non-zero mean drift.
Compared to the $\phi^4$ model,  a distinct mechanism for transport of energy over longer distances is available.  Such qualitative considerations
merely indicate that Fourier's law, while widely applicable, cannot be taken for granted. The precise value of the exponent $\alpha$ depends on $V$.
For example for the harmonic chain, $V(x) = x^2$, one would find $\alpha = 1$, because phonons propagate without collisions from one heat bath to the other.
Other integrable models, as the Toda lattice, exhibit the same behavior. Kinetic theory predicts $\alpha = 0.4$ \cite{lukkarinen2008}. Nonlinear fluctuating hydrodynamics
claims $\alpha = 0.33$, up to specified exceptions \cite{mendl2013,Spohn2014}, which seems to be widely confirmed \cite{das2014a}. Numerically difficulties arise because of possibly long cross-over scales \cite{das2014b,zhao2012,wang2013}. For example, for $V(x) = \tfrac{1}{2}x^2 - \tfrac{1}{3}x^3 + \tfrac{1}{4}x^4$ and a boundary temperature range of $0.1 -1.0$, even reaching $N =  65536$ the effective $\alpha$ varies between $0.09$ and $0.33$. It also depends on the choice of boundary conditions, either tied down or free  \cite{das2014b}.

For models in higher dimensions, the harmonic crystal exhibits $\alpha =1$ always. As soon as the interaction is nonlinear, no violation of Fourier's law
is known.  Possibly a systematic search is still missing.

Our models describe only elastic interactions, even then in a simplified form. Real systems have several channels of energy transport.
In addition the notion of perfect translation invariance is unrealistic. Disorder plays an important role, one example being isotope disorder.
This may suppress energy transport completely. But for harmonic crystals in three dimensions, in the presence of a quadratic pinning potential $U(q)$, isotope disorder  is responsible for the validity of Fourier's law \cite{kundu2010,chaudhuri2010}.
The interplay between nonlinearity and disorder is another important issue \cite{dhar2008b}. Thus Fourier's law is still an active and exciting area of research.


\begin{thebibliography}{99}

\bibitem{fourier} J. Fourier, Th\'{e}orie Analytique de la Chaleur. Chez Firmin  Didot, Paris, 1822\\
Translated as: The Analytical Theory of Heat, Cambridge University Press, London, 1878. 

\bibitem{bonetto2000} F. Bonetto, J.L. Lebowitz, and L. Rey-Bellet, \emph{ Fourier's law: a challenge to theorists}, in: Mathematical Physics 2000, A. Fokas, A. Grigoryan, T. Kibble, and B. Zegarlinski, eds., Imperial College Press, London, 2000, p. 128.

\bibitem{alder1970} B.J. Alder and T.E. Wainwright, \emph{ Decay of the velocity autocorrelation function}, Phys. Rev. A {\bf 1}, 18 (1970).

\bibitem{LLP2003} S. Lepri, R. Livi, and A. Politi, \emph{ Thermal conduction in classical low-dimensional lattices}, Phys. Rep. {\bf 377}, 1 (2003).

\bibitem{dhar2008}   A. Dhar, \emph{ Heat Transport in low-dimensional systems},
Adv. Phys. {\bf 57}, 457 (2008).

%
\bibitem{eckmann2000} J.-P. Eckmann and M. Hairer, \emph{ Non-equilibrium Statistical Mechanics of strongly anharmonic chains of oscillators}, Commun. Math. Phys. {\bf 212}, 105 (2000)

%
\bibitem{alder21970} B.J. Alder, D.M. Gass, and T.E. Wainwright, \emph{ Studies in molecular dynamics. VIII. The transport coefficients for a hard-sphere fluid}, J. Chem. Phys. {\bf 53}, 3813 (1970).

%
\bibitem{casati1984} G. Casati, J. Ford, F. Vivaldi, and W.M. Visscher,  \emph{ One-dimensional classical many-body system having a normal thermal conductivity}, Phys. Rev. Lett. {\bf 52}, 1861 (1984).

%
\bibitem{lepri1997} S. Lepri, R. Livi, and A. Politi, \emph{ Heat conduction in chains of nonlinear oscillators}, Phys. Rev. Lett. {\bf 78}, 1896 (1997).

\bibitem{Zhao00} B. Hu, B. Li, and H. Zhao, \emph{Heat conduction in one-dimensional nonintegrable systems}, Phys. Rev. E {\bf 61}, 3828 (2000).

\bibitem{Aoki02} K. Aoki, D. Kusnezov, \emph{Non-equilibrium Statistical Mechanics of classical lattice $\phi^4$ field theory}, Annals of Physics {\bf 295}, 50 (2002).

\bibitem{Zhao06} H. Zhao, \emph{Identifying diffusion processes in one-dimensional lattices in thermal equilibrium}, Phys. Rev. Lett. {\bf 96}, 140602 (2006).

\bibitem{Aoki06} K. Aoki, J. Lukkarinen and H. Spohn, \emph{Energy transport in weakly anharmonic chains}, J. Stat. Phys. {\bf 124}, 1105 (2006).

\bibitem{Lefevere06} R. Lefevere and A. Schenkel, \emph{Normal heat conductivity in a strongly pinned chain of anharmonic oscillators}, J. Stat. Mech.: Theory and Expt. {\bf L02001} (2006). 


\bibitem{Gendelman2000} O. V. Gendelman and A. V. Savin, \emph{Normal heat conductivity of the one-dimensional lattice with periodic potential of nearest-neighbor interaction},  Phys. Rev. Lett. {\bf 84}, 2381 (2000).

\bibitem{Das2015} S.G. Das and A. Dhar, \emph{Role of conserved quantities in normal heat transport in one dimension}, 	arXiv:1411.5247 (2015).

\bibitem{Iubini2016} S. Iubini, S. Lepri, R. Livi, and A. Politi, \emph{Coupled transport in rotor models}, New Journal of Physics, {\bf 18}, 083023 (2016).

%
\bibitem{das2019} A. Das, K. Damle, A. Dhar, D. A. Huse, M. Kulkarni, C.B. Mendl, and H. Spohn, \emph{ Nonlinear fluctuating hydrodynamics for the classical XXZ spin chain}, arXiv:1901.00024v1 (2019).

%
\bibitem{saito2010} K. Saito and A. Dhar, \emph{ Heat conduction in a three dimensional  anharmonic crystal}, Phys. Rev. Lett. {\bf 104}, 
040601 (2010).

\bibitem{Book} \emph{Lecture Notes in Physics vol 921 --- Thermal Transport in Low Dimensions: From Statistical Physics to Nanoscale Heat Transfer}, Ed. S. Lepri (Springer International Publishing 2016).

%
\bibitem{lukkarinen2008} J. Lukkarinen and H. Spohn, \emph{Anomalous energy transport in the FPU-$\beta$ chain}, Comm. Pure Appl. Math. {\bf 61}, 1753 (2008).

%
\bibitem{mendl2013} C. B. Mendl and H. Spohn, \emph{ Dynamic correlators of Fermi-Pasta-Ulam chains and nonlinear fluctuating hydrodynamics}, Phys. Rev. Lett. {\bf 111}, 230601 (2013).

%
\bibitem{Spohn2014}  H. Spohn, \emph{ Nonlinear fluctuating hydrodynamics for anharmonic chains},   J. Stat.Phys. {\bf 154},1191 (2014).


%
\bibitem{das2014a}  S.G. Das, A. Dhar, K. Saito, C.B. Mendl, and H. Spohn, \emph{ Numerical test of hydrodynamic fluctuation theory in the Fermi-Pasta-Ulam chain}, Phys. Rev. E {\bf 90}, 012124 (2014).

%
\bibitem{das2014b}  S.G. Das, A. Dhar, and O. Narayan, \emph{ Heat conduction in the $\alpha-\beta$ Fermi-Pasta-Ulam chain}, J. Stat. Phys. {\bf 154}, 204 (2014).

%
\bibitem{zhao2012} Y. Zhong, Y. Zhang, J. Wang, and H. Zhao, \emph{Normal heat conduction in one-dimensional momentum conserving lattices with asymmetric interactions},  Phys. Rev. E {\bf 85}, 060102(R) (2012).

%
\bibitem{wang2013} L. Wang, B. Hu, and B. Li, \emph{ Validity of Fourier's law in one-dimensional momentum-conserving lattices with asymmetric interparticle interactions}, Phys. Rev. E 88, 052112 (2013).

%
\bibitem{kundu2010}  A. Kundu, A. Chaudhuri, D. Roy, A. Dhar, J. L. Lebowitz, and H. Spohn, \emph{ Heat conduction and phonon localization in disordered harmonic  crystals}, Europhys. Lett. {\bf 90}, 40001 (2010).


%
\bibitem{chaudhuri2010} A. Chaudhuri, A. Kundu, D. Roy, A. Dhar, J.L. Lebowitz, and H. Spohn,  \emph{ Heat transport and phonon localization in mass-disordered harmonic crystals}, Phys. Rev. B {\bf 81}, 064301 (2010).


%
\bibitem{dhar2008b}  A. Dhar  and J. L. Lebowitz, \emph{ Effect of phonon-phonon interactions on localization}, Phys. Rev. Lett. {\bf 100}, 134301 (2008).



\end{thebibliography}
\end{document}